# Tips and Tricks to Improve CNN-based Chest X-ray Diagnosis: A Survey

Changhee HAN, Takayuki OKAMOTO, Koichi TAKEUCHI,

Dimitris KATSIOS, Andrey GRUSHNIKOV, Masaaki KOBAYASHI,

Antoine CHOPPIN, Yutaka KURASHINA, Yuki SHIMAHARA

LPIXEL Inc.

6F, Otemachi Building, 1-6-1, Otemachi, Chiyoda-ku, Tokyo, 100-0004, Japan

**Abstract:** Convolutional Neural Networks (CNNs) intrinsically requires large-scale data whereas Chest X-Ray (CXR) images tend to be data/annotation-scarce, leading to over-fitting. Therefore, based on our development experience and related work, this paper thoroughly introduces tricks to improve generalization in the CXR diagnosis: how to (*i*) leverage additional data, (*ii*) augment/distillate data, (*iii*) regularize training, and (*iv*) conduct efficient segmentation. As a development example based on such optimization techniques, we also feature LPIXEL's CNN-based CXR solution, EIRL Chest Nodule, which improved radiologists/non-radiologists' nodule detection sensitivity by 0.100/0.131, respectively, while maintaining specificity.

## 1. Introduction

Since many findings on Chest X-Ray (CXR), the world's most performed medical imaging test [1], are subtle or doubtful, CXR reading suffers from high inter-observer variability among even expert radiologists [2]. In this context, Convolutional Neural Networks (CNNs) have revolutionized CXR diagnosis (i.e., classification, regression, object detection, segmentation) [3-5]. However, the CNNs intrinsically requires ample data whereas the CXR images tend to be data/annotation-scarce, leading to over-fitting [6].

Therefore, as Fig.1 shows, this paper thoroughly introduces tricks to improve generalization in the CXR diagnosis: how to (*i*) leverage additional data, (*ii*) augment/distillate data, (*iii*) regularize training, and (*iv*) conduct efficient segmentation. A discussion on which specific CNN architectures to choose for Medical Imaging is out of scope in this paper, so refer to other surveys [7-9].

As a development example, we also feature LPIXEL's CNN-based CXR solution, EIRL Chest Nodule, which partially applies the introduced optimization techniques to empower doctors for more rapid and reliable nodule diagnosis. With the EIRL Chest Nodule assistance, radiologists/non-radiologists improved sensitivity by 0.100/0.131, respectively, while maintaining specificity.

## 2. Tricks to Improve CXR Diagnosis
### 2.1 Leveraging Public CXR Datasets

Since CNNs are data-hungry and obtaining large-scale CXR images is often unfeasible, it is essential to exploit publicly-available CXR images for pre-training, supervised learning, or semi-supervised learning. Currently, 4 large labeled open datasets exist: ChestXray-NIHCC [10] (~112,000 images); CheXpert [11] (~224,000 images); MIMIC-CXR [12] (~372,000 images); PadChest [13] (~160,000 images). For more detailed information on 20 public CXR datasets, refer to this review paper [14]. It should be noted that those datasets do not always provide definitive diagnosis information by Computed Tomography (CT) scans, which is associated with better prediction performance.

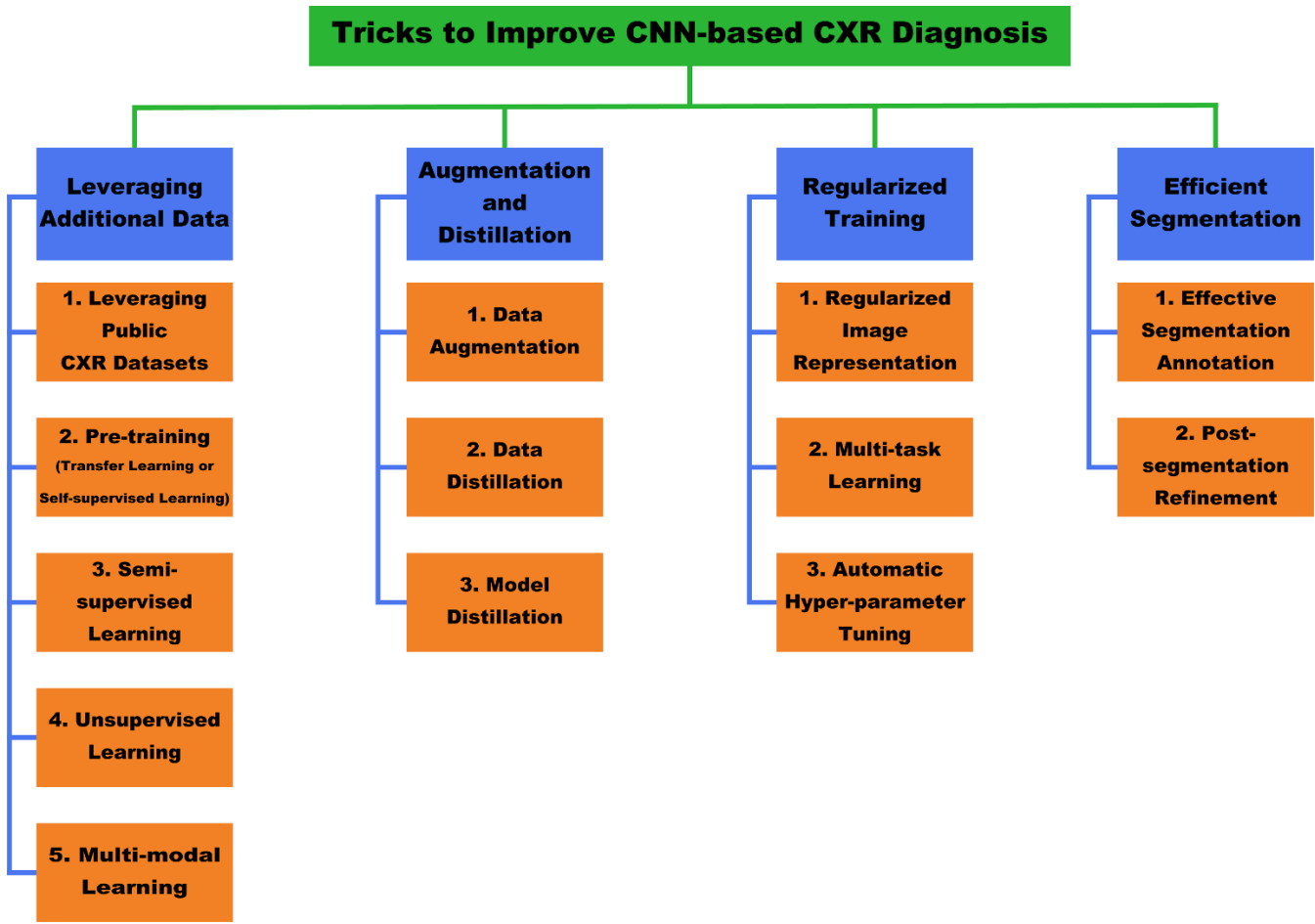

**Fig.1** Categorization of tricks to improve generalization in CXR diagnosis

**2.2 Pre-training**

Most dominant pre-training methods are (*i*) transfer learning, which uses labeled natural/medical images to obtain good initial weights, and (*ii*) self-supervised learning, which uses unlabeled medical images for initialization by solving auxiliary tasks based on input samples [6]. Generally, transfer learning on medical images (e.g., public CXR datasets) is ideal since the learned representation is strongly associated with the target medical task. However, due to the difficulty of data collection/annotation and unavailability of pre-trained models, transfer learning on natural images (e.g., ImageNet [15]/COCO [16]-based transfer learning) and self-supervised learning on medical images (e.g., Models Genesis [17], MoCo-CXR [18]) have been prevailing. Truncating final blocks of pre-trained models may significantly decrease parameters while statistically maintaining prediction performance on CXR images [19].

## 2.3 Semi-supervised Learning

Semi-supervised learning refers to training a model on both limited labeled/large-scale unlabeled (or pseudo-labeled) medical images to cut (especially segmentation) annotation cost [20]. Usually, effective semi-supervised learning requires at least thousands of labeled CXR images.

## 2.4 Unsupervised Learning

Unsupervised anomaly detection can discover various unseen abnormalities (e.g., rare disease, bleeding) without specifying disease types, relying on large-scale unannotated healthy medical images. Towards this, Generative Adversarial Networks (GANs) and (Variational) AutoEncoders reconstruct medical images to detect outliers either in the learned feature space or from high reconstruction loss [21].

## 2.5 Multi-modal Learning

Along with CXR images, concatenating patient data (e.g., age, gender, X-ray view position) to the (flattened) layer could improve prediction [22].

## 2.6 Data Augmentation

Data Augmentation (DA) plays a key role in improving generalization. Traditional DA adopts various intensity/geometric transformations (e.g., rotation, flipping, shearing, random resized cropping, changing contrast/sharpness); automated augmentation strategies (e.g., Randaugment [23]) can efficiently automate such DA by reducing a search space. In addition to the traditional DA, combining pairs of training images/labels (e.g., Mixup [24]) has shown promising performance, especially in medical image segmentation [25]. Conditional GAN-based DA also plays a big role in Medical Imaging, offering both interpolation/extrapolation effect [26, 27].

## 2.7 Data/Model Distillation

In Medical Imaging, Test-time DA and model ensembling assures robust model prediction [28], similar to dropout [29], which provides robustness in training network parameters. Training a model on both gray-scale/color images, respectively, and combining their results might also improve prediction [22].

## 2.8 Regularized Image Representation

Reducing a parameter space to a suitable subspace *via* regularized image representation helps avoid over-fitting on CXR images: multi-scale patch-based prediction [30] and resizing to a smaller image size simply reduces a search space; Conditional GAN-based denoising removes prediction-irrelevant noises while preserving image structure and details [31]; similarly, Conditional GAN-based bone suppression also increases the visibility of soft tissues by suppressing bones [32]; lung field detection isolates a lung region (i.e., region of interest) [4].

## 2.9 Multi-task Learning

As training with regularization, multi-task learning performs multiple tasks (e.g., classification, object detection, segmentation) using a single learned representation. In Medical Imaging, it typically refers to training a segmentation model with auxiliary heads, each for an individual classification task; urging the model to represent a classification-relevant (i.e., diagnosis-relevant) feature space tends to improve segmentation [33]. Since the prevalence of positive cases significantly differs across diseases, we need to address the data imbalance; wide-spread solutions are (*i*) under-sampling normal class, (*ii*) over-sampling rare class, (*iii*) Synthetic Minority Oversampling Technique [34], and (*iv*) weighted loss. In addition, combining multi-scale receptive fields helps capture diverse diseases varying in size [35].

**Table1** Physicians' reading performance with/without AI assistance.
AUC stands for Area Under the receiver operating characteristic Curve

|  | AUC | Sensitivity | Specificity |
|---|---|---|---|
| 9 Radiologists w/o AI Assistance | 0.717 ± 0.034 | 0.471 ± 0.611 | 0.964 ± 0.020 |
| 9 Non-radiology Physicians w/o AI Assistance | 0.700 ± 0.059 | 0.438 ± 0.112 | 0.963 ± 0.025 |
| 9 Radiologists w/ AI Assistance | 0.768 ± 0.021 | **0.571 ± 0.041** | 0.966 ± 0.026 |
| 9 Non-radiology Physicians w/ AI Assistance | **0.769 ± 0.031** | 0.569 ± 0.063 | **0.970 ± 0.022** |

**2.10 Automatic Hyper-parameter Tuning**

Since CNNs usually involve many hyper-parameters for tuning, manual searching is inefficient to handle the black box. Grid/random search also suffer from the curse of dimensionality and time inefficiency. Meanwhile, Bayesian optimization can effectively approximate the hyper-parameters from known samples (i.e., prior knowledge) on CXR images [3]. On top of the hyper-parameter tuning, reparametrization techniques allow for efficient large-batch training (e.g., batch normalization [36], Adaptive Gradient Clipping [37]).

**2.11 Effective Segmentation Annotation**

CXR reading suffers from high inter-observer variability. Therefore, computational ground truth prediction by estimating the annotator confusion leads to robust annotation [38]. Moreover, cost-effective annotation requires (*i*) active learning [39], which provides the annotators with samples to annotate that may improve generalization and (*ii*) interactive segmentation [40], which supports the annotators by propagating their modifications through the whole segmentation mask.

**2.12 Post-segmentation Refinement**

Post-segmentation refinement removes false positives and produces a smoother CXR segmentation: Kuzin *et al.* heuristically averaged cross-fold predictions using an optimized binarization threshold and a dilation technique [5]; Larrazabal *et al.* used a denoising autoencoder to obtain an anatomically plausible segmentation from the initial prediction [41].

**3. EIRL: Doctor's Anytime Assistant**

Japanese medical AI startup LPIXEL provides a variety of intelligent AI diagnostic solutions called EIRL Series to empower doctors for more rapid and reliable diagnosis. Specifically, EIRL Chest Nodule reliably detects nodules (between 5-30 mm) on CXR images by partially applying the numerous optimization techniques as mentioned in Section 2. Its version 1.8 adopts various DA methods (e.g., intensity/geometric augmentation) and post-processing methods (e.g., thresholding-based segmentation, isolated small area exclusion, lung field-based false positive reduction).

To validate the EIRL Chest Nodule's clinical efficacy, 18 physicians (9 radiologists/9 non-radiologists) took a reading test on 67 cases with nodules (76 nodules in total) and 253 cases without findings (Table1). With AI assistance, radiologists/non-radiologists improved sensitivity by 0.100/0.131, respectively, while maintaining specificity. Fig.2 shows example nodules found by a radiologist only with AI assistance, which implies the EIRL Chest Nodule's capacity to diagnose lung cancer at an early stage.

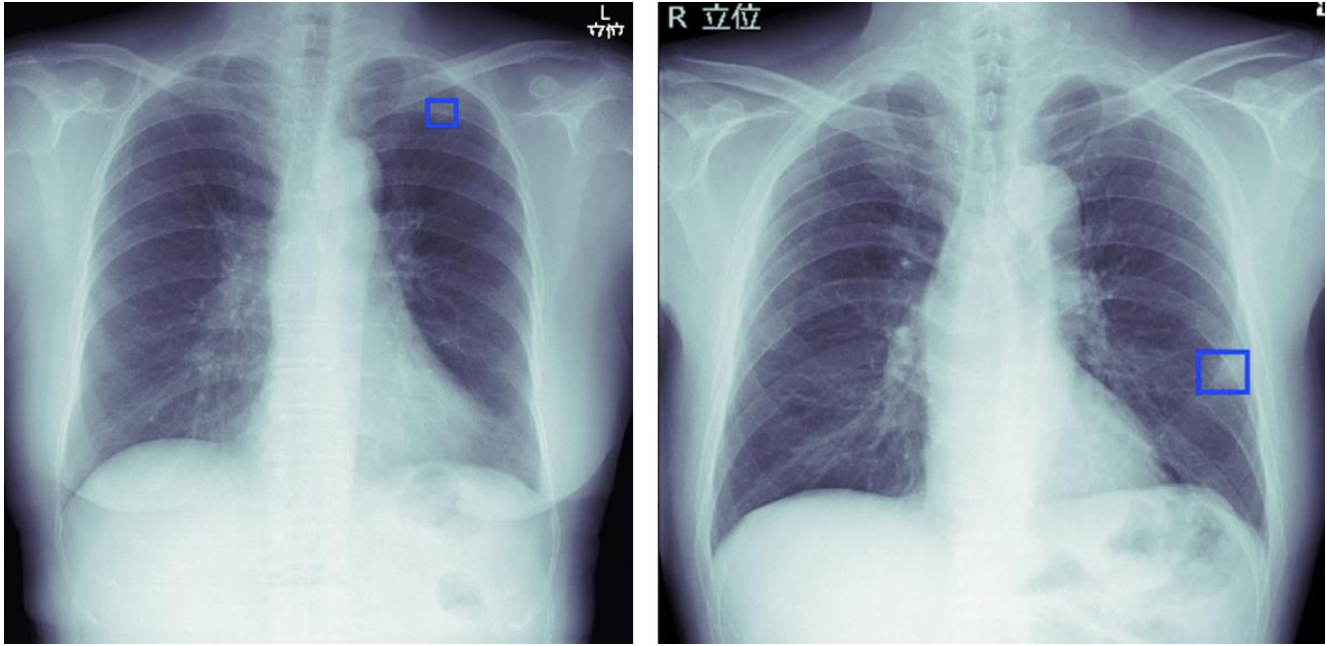

**Fig.2** Nodules found by a radiologist only with AI assistance

As a doctor's anytime assistant, the EIRL Chest Nodule has 4 features.

• **High-Quality Training Data**: It uses multi-institutional malignant nodule (i.e., primary lung cancer) data as training data, based on radiologists' agreements for CT findings.

• **Outstanding AI Model**: It adopts unique technology based on numerous state-of-the-art optimization techniques, including DA, transfer learning, and model ensembling.

• **Continuous Update**: It continues to add functions, cover more findings, and improve detection performance, whenever pharmaceutical affairs approval is given, by accumulating feedback/data from clinical environment and introducing cutting edge technology.

• **Excellent Compatibility**: It allows seamless and reliable integration/analysis with all major CXR scanners and Picture Archiving and Communication Systems (PACS).

## 4. Conclusion

Based on our development experience and related work, we thoroughly introduced various optimization tricks to improve CNN-based CXR diagnosis—many of them are generally applicable in Medical Imaging. Our CNN-based CXR solution, EIRL Chest Nodule, partially applies such techniques to empower doctors for more rapid and reliable nodule diagnosis. We plan to cover various findings on CXR while improving detection performance.